\documentclass[%
 reprint,
 amsmath,amssymb,
 aps,
nofootinbib
]{revtex4-2}

\usepackage{graphicx,amsmath}
\usepackage{dcolumn}
\usepackage{bm}
\usepackage[skins,theorems]{tcolorbox}
\usepackage{amssymb}
\usepackage{tablefootnote}
\usepackage{units}
\usepackage{threeparttable,lipsum}
\usepackage{mathtools}
\usepackage{hyperref}
\usepackage{dcolumn}
\usepackage{bm}

\tcbset{highlight math style={enhanced,
 colframe=cyan!10!gray,colback=gray!5!white,arc=4pt,boxrule=1pt,
  drop fuzzy shadow}}

  {\empheq[box=\tcbhighmath]{align}}
  {\endempheq}

\begin{document}

\preprint{APS/123-QED}

\title{Bayesian Estimation of the $S$ Factor and Thermonuclear Reaction Rate for $^{16}$O(p,$\gamma$)$^{17}$F}

\author{Christian Iliadis}
\author{Vimal Palanivelrajan}
\affiliation{%
 Department of Physics \& Astronomy, University of North Carolina at Chapel Hill, NC 27599-3255, USA\\
 Triangle Universities Nuclear Laboratory (TUNL), Durham, North Carolina 27708, USA
}%

\author{Rafael S. de Souza}
\affiliation{Key Laboratory for Research in Galaxies and Cosmology, Shanghai Astronomical Observatory,\\ Chinese Academy of Sciences, 80 Nandan Road, Shanghai 200030, China}

\date{\today}

\begin{abstract}
The $^{16}$O(p,$\gamma$)$^{17}$F reaction is the slowest hydrogen-burning process in the CNO mass region. Its thermonuclear rate sensitively impacts predictions of oxygen isotopic ratios in a number of astrophysical sites, including AGB stars. The reaction has been measured several times at low bombarding energies using a variety of techniques. The most recent evaluated experimental rates have a reported uncertainty of about 7.5\% below $1$~GK. However, the previous rate estimate represents a best guess only and was not based on rigorous statistical methods. We apply a Bayesian model to fit all reliable $^{16}$O(p,$\gamma$)$^{17}$F cross section data, and take into account independent contributions of statistical and systematic uncertainties. The nuclear reaction model employed is a single-particle potential model involving a Woods-Saxon potential for generating the radial bound state wave function. The model has three physical parameters, the radius and diffuseness of the Woods-Saxon potential, and the asymptotic normalization coefficients (ANCs) of the final bound state in $^{17}$F. We find that performing the Bayesian $S$ factor fit using ANCs as scaling parameters has a distinct advantage over adopting spectroscopic factors instead. Based on these results, we present the first statistically rigorous estimation of experimental $^{16}$O(p,$\gamma$)$^{17}$F reaction rates, with uncertainties ($\pm 4.2$\%) of about half the previously reported values.
\end{abstract}

\maketitle


\section{\label{sec:intro}Introduction}
The $^{16}$O(p,$\gamma$)$^{17}$F reaction ($Q$ $=$ $600.27 \pm 0.25$~keV \cite{Wang_2021}) is the slowest process among all proton-induced reactions in the CNO mass region \cite{ILIADIS2010b,Iliadis:2015ta}. The lowest-lying resonance is located at relatively high laboratory energy of $\approx$2.7~MeV \cite{Tilley1993}. Below this energy, the $^{16}$O(p,$\gamma$)$^{17}$F reaction is a prime example of the nonresonant direct radiative capture process, which assumes that the proton is captured via a single-step process into a final-state orbit outside a closed $^{16}$O core \cite{Rolfs1973,iliadis2004}. This reaction has been measured many times at low bombarding energies using a variety of techniques, including the activation method, in-beam detection of prompt $\gamma$ rays, and experiments in inverse kinematics. A comprehensive analysis of the most reliable data has been presented in Refs.~\cite{iliadis2008,Mohr2012}. The reported thermonuclear reaction rate uncertainty in Ref.~\cite{iliadis2008} is about 7.5\% at temperatures below $1$~GK. Knowledge of the rate at a few-percent uncertainty level is desirable because the $^{16}$O(p,$\gamma$)$^{17}$F reaction influences sensitively the $^{17}$O/$^{16}$O abundance ratio and, to a lesser degree, the $^{18}$O/$^{16}$O ratio. This information directly impacts the interpretation and paternity of oxygen isotopic ratios measured in presolar stardust grains \cite{2003MNRAS.340..763S,Lugaro2007,iliadis2008,10.3389/fspas.2020.607245}. 

Previous evaluations of the $^{16}$O(p,$\gamma$)$^{17}$F rate \cite{NACRE,iliadis2008,ILIADIS2010b} were performed with methods that were conventionally employed at the time. The data from different experiments were fitted independently because it was not clear, within the $\chi^2$ method used, how to perform a common fit across several different data sets. Also, it was neither clear how to treat independent contributions from statistical and systematic uncertainties, nor how to combine total cross section data in the analysis with data on individual transitions. For these reasons, a number of {\it ad hoc} assumptions were made by Ref.~\cite{NACRE,iliadis2008,ILIADIS2010b} that were not rigorous in a statistical sense.

The advent of thermonuclear rates based on hierarchical Bayesian models has improved this situation significantly. The method was first presented in Ref.~\cite{iliadis16} and subsequently applied to reactions of interest to Big Bang Nucleosynthesis (BBN). In the simplest cases, the Bayesian models employed either polynomial fitting functions or predictions from microscopic nuclear reaction models \cite{gomez17,deSouza:2019gi,Moscoso2021}. The method was extended in Refs.~\cite{deSouza:2019gf,de_Souza_2020} to implement a one-level R-matrix approximation into the Bayesian fitting. Here, we report on the implementation of a single-particle potential model into the Bayesian framework.

As will be seen below, the hierarchical Bayesian model solves a number of problems that plagued previous work (see, e.g., Ref.~\cite{iliadis2008}): (i) it allows for the straightforward implementation of the total cross section data of Ref.~\cite{Hester1958} and the single datum of Ref.~\cite{Becker1982} for the first-excited-state transition (both disregarded previously); (ii) it facilitates a combined fit of all data (results from different experiments were previously  fitted independently); (iii) it fully accounts for the independent contributions of statistical and systematic uncertainties  (which were previously combined into a single uncertainty); (iv) it makes no {\it ad hoc} assumptions on how to combine fits from different data sets or $\gamma$-ray transitions.

A primary goal of this work is to include the parameters of the single-particle potential model in the random sampling, i.e., we will be exploring the sensitivity of the $S$ factor to these quantities. It will be demonstrated that the uncertainty of the fitted $S$ factor is relatively large when the Bayesian analysis is performed using spectroscopic factors, but is significantly reduced when asymptotic normalization coefficients (ANCs) are employed instead. 

The data selection is discussed in Sec.~\ref{sec:data}. The nuclear reaction formalism is given in Sec.~\ref{sec:nuclear} and the hierarchical Bayesian model is presented in Sec.~\ref{sec:bayes}. Fits of the $S$ factor are found in Sec.~\ref{sec:results} and thermonuclear rates calculated in Sec.~\ref{sec:rates}. Section~\ref{sec:summary} provides a concluding summary.

\section{\label{sec:data}Data selection}
Consistent with previous Bayesian reaction rate estimates \cite{iliadis16,gomez17,deSouza:2019gf,deSouza:2019gi,de_Souza_2020,Moscoso2021}, we will consider only data for which statistical and systematic uncertainties can be estimated separately. This is the case for four data sets: Hester, Pixley and Lamb \cite{Hester1958}; Chow, Griffiths and Hall \cite{Chow1975}; Becker et al. \cite{Becker1982}; and Morlock et al. \cite{Morlock1997}. The first work \cite{Hester1958} only measured the total cross-section, while the third \cite{Becker1982} reported only the cross section for the transition into the first-excited state ($E_x$ $=$ $495.33 \pm 0.10$~keV \cite{Tilley1993}) at a single bombarding energy. The data of Refs.~\cite{Hester1958,Becker1982} were not taken into account in the previous $^{16}$O(p,$\gamma$)$^{17}$F rate evaluation \cite{iliadis2008}, because, at the time, it was neither clear how to fit the total cross section \cite{Hester1958} together with those for the individual transitions, nor how to reliably include a data set consisting of a single data point only \cite{Becker1982}. As will be seen in Sec.~\ref{sec:bayes}, all of these data can be rigorously included in a hierarchical Bayesian model. We discuss below the four data sets individually.

Hester, Pixley and Lamb \cite{Hester1958} measured the total cross-section of the $^{16}$O(p,$\gamma$)$^{17}$F reaction at six center-of-mass energies between $132$ and $160$~keV. These represent the lowest-energy data points among all the data sets taken into account in the present work. The reported statistical uncertainties range from 14\% to 40\% for the highest and lowest energy, respectively. The cross sections have been corrected using modern stopping powers, as discussed in \cite{iliadis2008}, and we adopt these corrected values in the present work. From their quoted uncertainties in the measured beam current (6\%), counter efficiency (7\%), and stopping power (10\%), we estimate a total systematic uncertainty of 14\%.

Chow, Griffiths and Hall \cite{Chow1975} measured the cross section for the transition to the ground state at four center-of-mass energies between $1288$ and $2404$~keV, and for the transition to the first-excited state at seven energies between $795$ and $2404$~keV. The statistical uncertainties range from 3\% to 12\%. The systematic effects in their measurement were discussed by Ref.~\cite{iliadis2008}, including $\gamma$-ray efficiency (3\%), escape peak detection (1\%), angular distribution correction (1\%), and effective bombarding energy (3\%). Consequently, we adopt a value of 5\% for the combined systematic uncertainty.

Becker et al. \cite{Becker1982} measured the $^{16}$O(p,$\gamma$)$^{17}$F reaction in inverse kinematics at a single center-of-mass energy of $853$~keV. Although not mentioned explicitly, their reported cross section refers to the transition to the first-excited state only. The statistical uncertainty amounts to 13\%. The main sources of systematic uncertainty arise from the strength of their adopted standard resonances in $^{19}$F(p,$\alpha_2$)$^{16}$O and the $\gamma$-ray efficiency in their extended gas target. We estimate an overall systematic uncertainty of 5\% for their reported cross section. 

Finally, Morlock et al. \cite{Morlock1997} reported $^{16}$O(p,$\gamma$)$^{17}$F cross sections below a center-of-mass energy of $3.5$~MeV. The lowest energy measured was $365$~keV for the ground state transition and $222$~keV for the first-excited state one. These data are presented in Fig.~3 of Ref.~\cite{Morlock1997}, which displays statistical uncertainties only. Subsequently, the Morlock et al. data \cite{Morlock1997} have been corrected by Ref.~\cite{iliadis2008} for coincidence summing, and these corrected data have been adopted for the present analysis. More detailed information regarding the cross section uncertainties of these data is given in the caption of Fig.~2.37 in Ref.~\cite{Morlock1997b}, which states ``...the statistical errors are with few exceptions between 1.5\% and 3\%. One has to add 10\% systematic uncertainty (scattering measurement, energy determinations, error propagation)...'' Consequently, in the present work, we adopt a systematic uncertainty of 10\%. The lowest-lying $^{16}$O(p,$\gamma$)$^{17}$F resonance is located near a center-of-mass energy of $2.5$~MeV, and, therefore, we only took their data points below an energy of $2.4$~MeV into account. The data at higher energies are irrelevant for stellar burning.

We note that the four measurements discussed above provide independent estimates of the $^{16}$O(p,$\gamma$)$^{17}$F cross section. In particular, only the work of Ref.~\cite{Hester1958} relied on stopping power corrections, while the other measurements of the direct capture cross section \cite{Chow1975,Becker1982,Morlock1997} were performed relative to the Rutherford scattering yield, thus obliviating the effects of target stoichiometry or stopping powers.

The cross sections, $\sigma$, discussed above were converted to astrophysical $S$-factors, defined by $S(E)$ $\equiv$ $\sigma(E) E e^{2\pi\eta}$, with $\eta$ denoting the Sommerfeld parameter. The experimental $S$-factors were then analyzed with our Bayesian model.

\section{\label{sec:nuclear}Nuclear reaction model}
The $^{16}$O(p,$\gamma$)$^{17}$F reaction cross section below $2.4$~MeV center-of-mass energy is considered as a standard case for the direct radiative capture (DC) model since the seminal works of Christy and Duck \cite{Christy1961} and Rolfs \cite{Rolfs1973}. Subsequent analyses using the direct capture model for $^{16}$O(p,$\gamma$)$^{17}$F can be found in Refs.~\cite{brune1996,Gagliardi1999,iliadis2004,iliadis2008}, and references therein. The study of Ref.~\cite{iliadis2008} demonstrated that the potential model and the R matrix model provide nearly identical data fits at low energies. We will adopt in the present work a single-particle potential model, as discussed below.

The potential model assumes a single-step process, where the proton is directly captured, without the formation of a compound nucleus, into a final bound state with the emission of a photon. The dominant E1 contribution to the theoretical (p,$\gamma$) cross section (in $\mu$b) for capture from an initial scattering state with orbital angular momentum $\ell_i$ to a final bound state with orbital angular momentum $\ell_{f}$ and the principal quantum number $n$ (i.e., the number of wave-function nodes), is given by \cite{Rolfs1973}
\begin{align}
\label{eq:sp1}
\sigma^{DC}_{sp}(E1, n, \ell_{i}, \ell_{f}) = 0.0716 \mu^{\frac{3}{2}}\left (\frac{Z_{p}}
         {M_{p}} - \frac{Z_{t}}{M_{t}} \right)^{2} \frac{E_{\gamma}^{3}}
         {E^{\frac{3}{2}}} \times \notag 
         \\ 
         \frac{(2J_{f}+1)(2\ell_{i}+1)}
         {(2j_{p}+1)(2j_{t}+1)(2\ell_{f}+1)} (\ell_{i} 0 1 0 | \ell_{f} 0)^{2}
         R_{n \ell_{i} 1 \ell_{f}}^{2} 
\end{align}
\begin{equation}
\label{eq:sp2}
    R_{n \ell_{i} 1 \ell_{f}} =
         \int_{0}^{\infty} u_{s}(r) \mathcal{O}_{E1}(r) u_{b}(r) dr 
\end{equation}
where $\mu$ is the reduced mass, $Z_t$, $Z_p$ and $M_t$, $M_p$ are the charges and masses (in amu), respectively, of target and projectile; $j_{p}$, $j_{t}$, $J_{f}$ are the spins of projectile, target and final state, respectively; $E$ and $E_\gamma$ are the center-of-mass energy and the emitted $\gamma$-ray energy, respectively; $\mathcal{O}_{E1}$ is the radial part of the E1 multipole operator; and $u_{s}$ and $u_{b}$ are the radial wave functions of the initial scattering state and final bound state, respectively, where $u_b(r=0)$ $=$ $0$ and $\int_0^\infty{u_b^2 dr}$ $=$ $1$. We disregarded any M1 and E2 contributions, which amount to less than $0.1$\% compared to the dominant E1 $S$ factor \cite{Rolfs1973,Des1999}. 

Bound-state wave functions were generated by using a potential consisting of a Woods-Saxon term and a Coulomb term, given by
\begin{equation}
    V(r) = \frac{-V_0}{1 + e^{(r-R)/a}} + V_C(r) 
\end{equation}
where $R$ $=$ $r_0 A_t^{1/3}$ and $a$ are the Woods-Saxon potential radius and  diffuseness, respectively; $A_t$ is the mass number of the target nucleus; $V_C$ corresponds to a uniformly charged sphere of radius $R$. The well depth, $V_0$, was chosen to reproduce the binding energy of the final state. Several works (see summary in Table~I of Ref.~\cite{iliadis2004}) have employed bound-state square-well potentials instead of Woods-Saxon potentials. However, Refs.~ \cite{Wiescher1980,Powell1999} found that the adoption of square-well potentials over-predicts the calculated single-particle direct capture cross sections by up to a factor of $3$.

Scattering-state wave functions were computed using a hard-sphere nuclear potential, which gives similar results as a zero-energy nuclear potential at the low energies explored here \cite{iliadis2004}. The insensitivity of the $^{16}$O(p,$\gamma$)$^{17}$F cross section to the choice of scattering potential at low energies has also been reported in Ref.~\cite{Akram2001}. 
The hard-sphere nuclear potential radius was set equal to the Woods-Saxon radius, $R$, of the bound state potential. The radial integration in Eq.~(\ref{eq:sp2}) was extended to $500$~fm, because the integrand has a maximum located far beyond the nuclear radius at the lowest center-of-mass energies explored here. For the same reason, we used the exact expression for the radial part of the E1 operator, $\mathcal{O}_{E1}$, instead of its long-wavelength approximation.

For a zero-spin target nucleus, such as $^{16}$O, the direct capture to a specific final state proceeds via a unique orbital angular momentum, $\ell_f$, but may involve several values of $\ell_i$. In this case, the direct capture cross section is given by an incoherent sum, 
\begin{equation}\label{eq:dc}
\sigma^{DC} = C^{2}S_{\ell_f}~\sum_{\ell_{i}} 
               \sigma^{DC}_{sp}(E1, n, \ell_{i}, \ell_{f})
\end{equation}
with S$_{\ell_f}$ and $C$ denoting the spectroscopic factor and isospin Clebsch-Gordan coefficient, respectively. For $^{16}$O $+$ $p$, we have $C^2$ $=$ $1$. To avoid confusion with other symbols, we will not mention further the isospin Clebsch-Gordan coefficient.

For the $^{16}$O(p,$\gamma$)$^{17}F$ reaction data considered in the present work, the population of the $^{17}$F ground state, $DC$ $\rightarrow$ $0$~keV ($J_f$ $=$ $5/2^+$), proceeds predominantly via $E1$ radiation and orbital angular momenta of $\ell_i$ $=$ $1$, $3$ and $\ell_f$ $=$ $2$, while the transition to the first excited state, $DC$ $\rightarrow$ $495$~keV ($J_f$ $=$ $1/2^+$), proceeds via E1 radiation and angular momenta of $\ell_i$ $=$ $1$ and $\ell_f$ $=$ $0$. We assume that the proton is transferred into the $1d_{5/2}$ and $2s_{1/2}$ shell-model orbitals for the transition to the ground and first-excited state, respectively. In the following, we will label the spectroscopic factor (or later the ANC) for a given final level by the orbital angular momentum of the bound state alone, and suppress other quantum numbers (such as total spin). Hence, $S_{gs}$ $\equiv$ $S_{\ell_f = 2}$ and $S_{fes}$ $\equiv$ $S_{\ell_f = 0}$ for the transition to the ground and first-excited state, respectively. 

The calculated single-particle cross section, $\sigma^{DC}_{sp}$, will depend strongly on the adopted choice of the Woods-Saxon potential radius parameter, $r_0$, and diffuseness, $a$. This means that the spectroscopic factor, $S_{\ell_f}$, which is derived from a fit to experimental data, will also be sensitive to these very same parameters. We will demonstrate below the difficulty arising when the data are fitted in terms of the spectroscopic factor using Eq.~(\ref{eq:dc}).

Previous works have demonstrated \cite{Rolfs1973,Morlock1997} that, for the $^{16}$O(p,$\gamma$)$^{17}$F reaction at low energies, the integrand in Eq.~(\ref{eq:sp2}) peaks far outside the nuclear radius. For such a peripheral reaction, the single-particle radial bound state wave function is asymptotically given by \cite{Akram2001}
\begin{equation}\label{eq:whitt}
u_{b,\ell_f}(r) \rightarrow b_{\ell_f} W_{-\eta, \ell_f+1/2}(2 \kappa r)
\end{equation}
where $b_{\ell_f}$ is the single-particle asymptotic normalization coefficient (spANC) and $W$ is the Whittaker function \cite{Hebbard1963}; $\kappa$ is the bound state wave number, with $\kappa^2$ = $2 \mu E_b / \hbar^2$, where $\mu$ is the reduced mass and $E_b$ $=$ $Q - E_x$ the binding energy of the final state; $\eta$ $=$ $ e Z_p Z_t \mu / (\kappa \hbar^2 )$ is the bound state Coulomb parameter. 

In a microscopic nuclear model, the capture cross section can be described in terms of the overlap, $I_B$, of the $^{16}$O, $^{17}$F, and $p$ bound-state wave functions and a many-body wave function for the relative motion. Assuming a single-particle model, the radial dependence of the overlap function, which represents the projection of the bound final $^{17}$F state onto the product bound state wave functions of $^{16}$O and the proton, can be approximated by
\begin{equation}\label{eq:overlap}
I_B(r) \approx \sqrt{S_{\ell_f}} u_{b,\ell_f}(r)
\end{equation}
At large distances between target and projectile, the consequences of the complicated many-body effects will vanish and the radial dependence of the overlap function becomes asymptotically
\begin{equation}\label{eq:anc}
I_B(r) \rightarrow C_{\ell_f} W_{-\eta, \ell_f+1/2}(2 \kappa r)
\end{equation}
where $C_{\ell_f}$ is the asymptotic normalization coefficient (ANC). Comparison of Eqs.~(\ref{eq:whitt}) $-$ (\ref{eq:anc}) yields the relationship between $S_{\ell_f}$, $C_{\ell_f}$, and $b_{\ell_f}$ \cite{Akram2001}
\begin{equation}\label{eq:relation}
S_{\ell_f} = \frac{C^2_{\ell_f}}{b^2_{\ell_f}}
\end{equation}
In this approach, $C_{\ell_f}^2$ appears as an observable quantity, while both $S_{n \ell_f}$ and $b_{\ell_f}^2$ are derived quantities that depend strongly on the parameters of the assumed single-particle potential model. 
 
The nuclear model discussed above has three parameters: (i) the ANC, $C_{\ell_f}^2$ (or the spectroscopic factor, $S_{\ell_f}$) specified by the final $^{17}$F state; (ii) the radius parameter, $r_0$, of the Woods-Saxon, Coulomb, and hard-sphere potentials; and (iii) the diffuseness, $a$. We will treat these three as adjustable parameters in the fitting. It can be seen from Eqs.~(\ref{eq:dc}) and (\ref{eq:relation}) that $C_{\ell_f}^2$ or $S_{\ell_f}$ act as multiplicative scaling factors of the computed single-particle potential cross section (or $S$ factor) for a given transition. 

Based on the above discussion, we expect that varying $r_0$ and $a$ will impact the extracted value of $S_{\ell_f}$ significantly. When the fit is performed instead using $C_{\ell_f}^2$, this dependence should be much lessened, but may not entirely disappear because of the approximation in Eq.~(\ref{eq:overlap}). In either case, we expect that the fit will not constrain $r_0$ or $a$. These effects will be explored in Sec.~\ref{sec:results}.


\section{\label{sec:bayes}Bayesian model}
The hierarchical Bayesian model constructed for the present work is similar to those presented in Refs.~\cite{iliadis16,gomez17,deSouza:2019gf,deSouza:2019gi,de_Souza_2020,
Moscoso2021}. The model takes all relevant effects impacting the measured data into account. The inference framework is based on Bayes' theorem \cite{2017bmad}
\begin{equation}
\label{eq:Bayes}
    p(\theta|y) = \frac{\mathcal{L}(y|\theta)\pi(\theta)}{\int \mathcal{L}(y|\theta)\pi(\theta)d\theta}
\end{equation}
where the data are represented by the vector $y$ and the complete set of model parameters is denoted by the vector $\theta$. All factors in Eq.~(\ref{eq:Bayes}) represent probability densities: $\mathcal{L}(y|\theta)$ is the likelihood, i.e., the probability that the data, $y$, were obtained assuming given values of the model parameters, $\theta$; $\pi(\theta)$ is the prior, which represents our state of knowledge about each parameter before seeing the data; the product of likelihood and prior defines the posterior, $p(\theta|y)$, i.e., the probability of obtaining the values of a specific set of model parameters given the data. Bayes' theorem implies that the posterior represents an update of our prior state of knowledge regarding the model parameters once new data become available. The denominator, called evidence, is a normalization factor and is not important for the discussion of the present work. 

When the experimental $S$ factor is subject to statistical uncertainties only, $\sigma_{stat}$, the likelihood is given by
\begin{equation}
\label{eq:likelihood} 
    \mathcal{L}(S^{exp}|\theta)=\prod_{i=1}^N \frac{1}{\sigma_{stat,i}\sqrt{2\pi}}
    e^{-\frac{[S_i^{exp} - S(\theta)_i]^2}{2\sigma^2_{stat,i}}}
\end{equation}
where $S(\theta)_i$ $=$ $S_{i}^{DC}$ is the theoretical $S$ factor, as predicted by direct capture theory (Eqs.~(\ref{eq:dc}) and (\ref{eq:relation})), while the product runs over all the data points, labeled by the index $i$. The likelihood represents a product of normal distributions, each with a mean of $S(\theta)_i$ and a standard deviation of $\sigma_{stat,i}$, with the latter quantity given by the experimental statistical uncertainty of datum $i$. In symbolic notation, we can express Eq.~(\ref{eq:likelihood}) as
\begin{equation}
\label{eq:Si}
S_i^{exp} \sim N(S(\theta)_i,\sigma^2_{stat,i})
\end{equation}
where ``$N$'' denotes a normal (Gaussian) probability density and the symbol ``$\sim$'' stands for ``has the probability distribution of.''

According to Eqs.~(\ref{eq:dc}) and (\ref{eq:relation}), the direct capture $S$ factor, $S^{DC}$, for a given transition (i.e., to the ground or first-excited state) at the center-of-mass energy, $E$, has the form
\begin{equation}
\label{eq:theory}
S^{DC}(E) = S_{\ell_f}  S_{sp}^{DC}(E, r_0, a) = C^2_{\ell_f} \frac{S_{sp}^{DC}(E, r_0, a)}{b^2_{\ell_f}(r_0, a)}
\end{equation}
where $S_{sp}^{DC}$ denotes the single-particle direct capture $S$ factor of Eq.~(\ref{eq:dc}). When the data analysis is performed using the ANC as scaling factor to fit the data, the ratio $S_{sp}^{DC}/b^2_{\ell_f}$ is computed by the nuclear model. Alternatively, when the fit is scaled using the spectroscopic factor, the nuclear model calculates the quantity $S_{sp}^{DC}$. Both $S_{sp}^{DC}$ and $b^2_{\ell_f}$ depend explicitly on the parameters $r_0$ and $a$ of the Woods-Saxon potential. The total $^{16}$O(p,$\gamma$)$^{17}$F $S$ factor is given by the incoherent sum over the individual transitions. The results quoted in Sec.~\ref{sec:results} will be based on adopting $C^2_{\ell_f}$ as the scaling factor. We will briefly discuss the outcomes of tests performed when adopting instead the spectroscopic factor.

Each model parameter requires a prior. Since one goal of the present work was to explore the impact of Woods-Saxon potential parameter variations over wide ranges, we adopted weakly-informative priors for the radius parameter, $r_0$, and diffuseness, $a$,
\begin{align}
\label{eq:prior1}
r_0 \sim U(0.9~\mathrm{fm}, 1.5~\mathrm{fm}) \\
a \sim U(0.5~\mathrm{fm}, 0.7~\mathrm{fm}) \label{eq:prior1b}
\end{align}
where ``$U$'' denotes a uniform (i.e., constant) probability density between the stated boundaries. Outside the ranges given above, the numerical integration does not have a solution with the required number of nodes in the bound state wave function, $u_b(r)$, when the depth of the Woods-Saxon potential is adjusted to reproduce the binding energy of the experimental level.

Previous work \cite{baye1998,Gagliardi1999,iliadis2008,Azuma2010} has established ANC values near $C^2_{gs}$ $\approx$ $1$ and $C^2_{fes}$ $\approx$ $7000$ for the transitions to the ground and first-excited states, respectively (see Sec.~\ref{sec:results}). Our goal is to predict ANCs based on directly measured $^{16}$O(p,$\gamma$)$^{17}$F data alone. In particular, we do not want to restrict the sampling ranges of the ANCs based on prior information from either transfer measurements or theoretical model calculations. For this reason, we will adopt non-informative priors 
\begin{align}
\label{eq:prior2}
C^2_{gs} \sim T(0,\infty)N(0, [10^6~\mathrm{fm^{-1}}]^2) \\
C^2_{fes} \sim T(0,\infty)N(0, [10^6~\mathrm{fm^{-1}}]^2) 
\end{align}
where ``$N$'' denotes a truncated normal density centered at zero with a large standard deviation, multiplied by a truncation function, ``$T$'', which suppresses samples outside of the interval defined by its arguments. 

%
%

Following Ref.~\cite{iliadis16}, we will incorporate into our model the systematic uncertainty in a given experiment as an informative, lognormal prior, given by
\begin{equation}
\label{eq:lognor}
 \pi(f) = \frac{1}{\ln (f.u.) \sqrt{2\pi}f}e^{-\frac{[\ln f]^2}{2[\ln (f.u.)]^2}}
\end{equation}
or in symbolic notation
\begin{equation}
\label{eq:lognor2}
f \sim LN(0, [\ln (f.u.)]^2)
\end{equation}
where ``$LN$'' denotes a lognormal probability density. In Eqs.~(\ref{eq:lognor}) and (\ref{eq:lognor2}), the lognormal location parameter, $\mu$, is equal to zero, i.e., $\mu$ $=$ $\ln x_{med}$ $=$ $0$. In other words, we are assuming that the median value of the lognormal distribution, $x_{med}$, equals unity. If this would not be the case, we could have corrected the data for the systematic effect. The lognormal spread parameter, $\sigma$, in Eqs.~(\ref{eq:lognor}) and (\ref{eq:lognor2}) is chosen as $\sigma$ $=$ $\ln (f.u.)$, where $f.u.$ denotes the factor uncertainty, which is given by the systematic uncertainty. For example, we pointed out in Sec.~\ref{sec:data} that the data of Hester, Pixley and Lamb \cite{Hester1958} have a systematic uncertainty of 14\%. In this case, we have chosen $\sigma$ $=$ $\ln (f.u.)$ $=$ $\ln (1.14)$. The posterior of the ``normalization factor,'' $f$, will describe by how much a specific data set deviates from the best-fit line.

In conventional $\chi^2$ fitting, normalization factors are viewed as a systematic shift in the data. In Bayesian inference, on the other hand, the reported data are never modified. Instead, the true (but unknown) S-factor is multiplied by the normalization factor, $f$. This means that, during the fitting, each data set pulls on the true S-factor curve with a strength inversely proportional to the systematic uncertainty: a data set with a small systematic uncertainty will pull the true S-factor curve more strongly toward it compared to a set with a large systematic uncertainty. This ``pulling'' is independent of the data set size: disregarding statistical uncertainties for a moment, a set consisting of a single datum will have the same weight in the fitting as one containing many data points, if both sets are described by the same factor uncertainty, $f.u.$  

In previous work (e.g., Ref.~\cite{deSouza:2019gf}), the effect of ``extrinsic'' scatter was explicitly included in the Bayesian model to account for an additional and unreported source of statistical uncertainty. For the case of $^{16}$O(p,$\gamma$)$^{17}$F, we find no compelling reason to include this effect. Neither the data points of Hester, Pixley and Lamb \cite{Hester1958}, nor those of Chow, Griffiths and Hall \cite{Chow1975} exhibit a scatter inconsistent with their reported uncertainties. The data of Morlock et al. \cite{Morlock1997}, both for the transition to the ground and first-excited state, indeed reveal a scatter that clearly exceeds their reported statistical and systematic uncertainties. But this affects only the data points at their lowest measured energies, which also appear to be systematically too high (see below). Therefore, it would be inappropriate to include this effect as ``extrinsic scatter'' in the Bayesian model. To test whether the observed low-energy scatter in the data of Ref.~\cite{Morlock1997} influences any of our results, we implemented a robust algorithm into the Bayesian model that accounts for outliers \cite{andreon2015bayesian,iliadis16}. It was found that the observed low-energy scatter does not impact the fit.


Our complete Bayesian model for a given transition, either to the ground or first-excited state, can be summarized below in symbolic notation:\\ 
\begin{align}
\label{eq:model}
    & \textrm{Nuclear reaction model:} \notag \\
    & \indent S(\theta)_i = C^2_{\ell_f} \frac{S_{sp}^{DC}(E, r_0, a)}{b^2_{\ell_f}(r_0, a)} \\  
    & \textrm{Parameters:} \notag \\
    & \indent \theta \equiv ( C^2_{\ell_f}, r_0, a ) \\     
    & \textrm{Likelihoods:}\notag \\ 
    & \indent S^{\prime}_{i,j} = f_{j} \times S(\theta)_{i} \\
    & \indent S^{exp}_{i,j} \sim N(S^{\prime}_{i,j}, \sigma_{stat,i}^2) \\
    & \textrm{Priors:}\notag \\ 
    & \indent C^2_{\ell_f} \sim T(0,\infty)N(0, [10^6~\mathrm{fm^{-1}}]^2) \\
    & \indent r_0 \sim U(0.9~\mathrm{fm}, 1.5~\mathrm{fm}) 	\\
    & \indent a \sim U(0.5~\mathrm{fm}, 0.7~\mathrm{fm})  \\
    & \indent f_j \sim LN(0, [\ln (f.u.)_j]^2) \label{eq:last}
\end{align}
The index $i$ labels individual data points, and $j$ denotes the experiment. 

As mentioned earlier, Hester, Pixley and Lamb \cite{Hester1958} only measured the total $S$ factor. These data points were simply described in our Bayesian model by the sum of the ground and first-excited-state transitions. Therefore, they were analyzed simultaneously with all other data sets and thereby constrain the strengths of the transitions to the individual levels. At the same time, the fitting correlations between the two partial $S$ factors are fully contained in our results. We note that in the fitting we did not take into account the $\pm2$~keV beam energy uncertainties reported in Ref.~\cite{Hester1958}. Their effect on the fit will be much smaller than the uncertainties of the reported cross sections, as can be seen from their Fig.~4.

For the analysis of the Bayesian model, we employed the program \texttt{JAGS} (“Just Another Gibbs Sampler”) using Markov chain Monte Carlo (MCMC) sampling \citep{Plummer03jags:a}. Specifically, we will employ the \texttt{rjags} package that works directly with \texttt{JAGS} within the R language \citep{Rcitation}. Running a \texttt{JAGS} model refers to generating random samples from the posterior distribution of model parameters. This involves the definition of the model, likelihood, and priors, as well as the initialization, adaptation, and monitoring of the Markov chains.

The MCMC sampling will provide the posteriors of all parameters. For each simulation, we computed three MCMC chains, where each had a length of $3 \times 10^6$ steps after disregarding the burn-in samples ($10^5$ steps for each chain). This ensured that the chains reached equilibrium, that effective chain lengths for all parameters exceeded $10000$, and that Monte Carlo fluctuations were negligible compared to statistical and systematic uncertainties.

\section{\label{sec:results}Results}
\begin{figure}[ht]
\includegraphics[width=1.0\columnwidth]{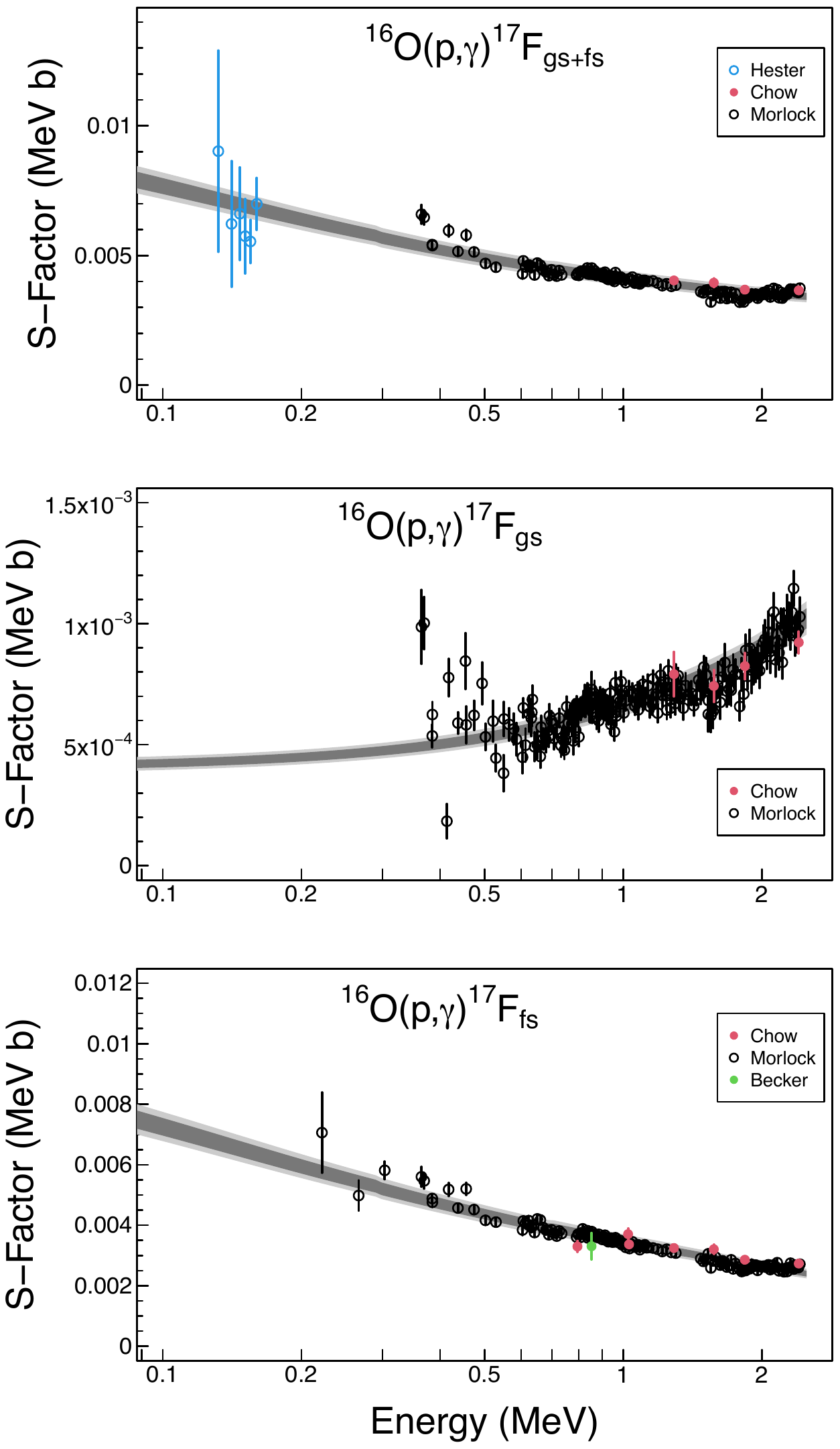}
\caption{\label{fig:fits} 
Results of Bayesian fits to experimental $^{16}$O(p,$\gamma$)$^{17}$F $S$ factors \cite{Hester1958,Chow1975,Becker1982,Morlock1997}: (Top) Total $S$ factor; (Middle) $S$ factor for the transition to the $^{17}$F ground state; (Bottom) $S$ factor for the transition to the $^{17}$F first-excited state. Becker et al. \cite{Becker1982} measured only the $S$ factor for the first-excited state transition at a single energy (green datum in the bottom panel). Hester, Pixley and Lamb \cite{Hester1958} measured only the total $S$ factor (blue data points in the top panel). The dark and light shaded regions represent coverage probabilities of 68\% and 95\%, respectively. Notice the logarithmic scale on the abscissa. 
}
\end{figure}

The resulting $S$-factor fits, adopting the model presented in Eqs.~(\ref{eq:model}) $-$ (\ref{eq:last}), are presented in Fig.~\ref{fig:fits}. The top panel corresponds to the sum of the ground and first-excited-state transitions, while the middle and bottom ones depict the fits to the individual transitions \footnote{The negative slope of the $S$ factor for the transition to the first-excited state has been variously explained by a pole in the $S$ factor \cite{PhysRevC.58.579} or a supposed halo property of the first-excited state in $^{17}$F \cite{Morlock1997}. The correct explanation was given by Ref.~\cite{MUKHAMEDZHANOV2002437}, who demonstrated that, for weakly-bound final states, the upturn of the $S$ factor towards lower energies arises naturally from the interplay of the Whittaker function corresponding to the bound state and the regular Coulomb wave function associated with the scattering state.}. As pointed out in Sec.~\ref{sec:data}, only Chow, Griffiths and Hall \cite{Chow1975} (red full circles) and Morlock et al. \cite{Morlock1997} (black open circles) measured cross sections for both transitions separately. The single data point of Becker et al. \cite{Becker1982} (green full circle) was measured for the first-excited-state transition and, thus, appears only in the bottom panel. Hester, Pixley and Lamb \cite{Hester1958} (blue open circles) measured the total cross section and their data appear in the top panel only. The dark and light shaded areas represent our $S$ factor predictions for 68\% and 95\% coverage probabilities, respectively. 

Numerical results are listed in Table~\ref{tab:results1}. For the total (i.e., the sum of ground and first-excited state transition) zero-energy S factor, we find a value of $S(0)$ $=$ $0.01092$~MeVb~($\pm$4.0\%), where the quoted uncertainties are derived from the 16, 50, and 84 percentiles. Our result agrees with that of Gagliardi et al. \cite{Gagliardi1999}, although their value has a significantly larger uncertainty (10\%). The latter work determined the ANCs of the ground and first-excited state by measuring the $^{16}$O($^3$He,d)$^{17}$F transfer reaction as a test case for the indirect estimation of the low-energy $^{16}$O(p,$\gamma$)$^{17}$F S factor. Our best estimate for the total S factor at $E$ $=$ $0.090$~MeV is $S(0.090)$ $=$ $0.00786$~MeVb~ ($\pm$4.0\%). This agrees with the average value found by Iliadis et al. \cite{iliadis2008} for R-matrix and potential-model fits to the data of Refs.~\cite{Chow1975,Morlock1997}, but their uncertainty (7.5\%) was not determined rigorously in a statistical sense, as discussed in Sec.~\ref{sec:intro}. Our value also agrees with the result obtained in the R-matrix analysis of the Morlock et al. \cite{Morlock1997} data by Azuma et al. \cite{Azuma2010}, but no uncertainties were reported in the latter work. For future reference, we provide in Table~\ref{tab:results3} numerical values of our recommended $S$ factor for a grid of energies.
\begin{table*}[]
\begin{center}
\caption{Summary of ANCs and $S$ factors for $^{16}$O(p,$\gamma$)$^{17}$F.}
\label{tab:results1}
\begin{ruledtabular}
\begin{tabular}{lrrcccc}
 & \multicolumn{2}{c}{Present} & \multicolumn{4}{c}{Previous} \\  \cline{2-3} \cline{4-7} 
                                        & \footnotemark[1]         &  \footnotemark[2]        &  Baye\footnotemark[3] \cite{baye1998}   &   Gagliardi\footnotemark[4] \cite{Gagliardi1999}  & Iliadis \cite{iliadis2008} & Azuma\footnotemark[7] \cite{Azuma2010} \\
\hline
$C^2_{gs}$~(fm$^{-1}$)                  &  {\bf 1.115~($\pm$4.0\%)}    &  $1.006$~($\pm$4.0\%)    &  $1.19$, $0.947$        &  $1.08$~($\pm$9.3\%)              &  $1.34$\footnotemark[5]    &  $1.10$    \\
$C^2_{fes}$~(fm$^{-1}$)                 &  {\bf 7063~($\pm$4.0\%)}     &  $6759$~($\pm$4.0\%)     &  $8306$, $7468$         &  $6490$~($\pm$10\%)               &  $6667$\footnotemark[5]    &  $6512$    \\
$S_{gs+efs}(0~\mathrm{MeV})$~(MeVb)     &  {\bf 0.01092~($\pm$4.0\%)}  &  $0.01040$~($\pm$4.0\%)  &                         &  0.0102~($\pm$10\%)               &                            &            \\
$S_{gs+efs}(0.09~\mathrm{MeV})$~(MeVb)  &  {\bf 0.00786~($\pm$4.0\%)}  &  $0.00751$~($\pm$4.0\%)  &                         &                                   &  $0.0076$~($\pm$7.2\%)\footnotemark[6]     &  $0.00807$ \\
\end{tabular}
\end{ruledtabular}
\footnotetext[1] {\footnotesize Results of fit to all $^{16}$O(p,$\gamma$)$^{17}$F data adopting broad priors. See text.}
\footnotetext[2] {\footnotesize Results of fit to all $^{16}$O(p,$\gamma$)$^{17}$F data for fixed values of $r_0$ $=$ $1.25$~fm and $a$ $=$ $0.65$~fm. See text. }
\footnotetext[3] {\footnotesize Values calculated using a microscopic nuclear model. }
\footnotetext[4] {\footnotesize Results of fit to $^{16}$O($^3$He,d)$^{17}$F data. }
\footnotetext[5] {\footnotesize Results of R-matrix fits to the $^{16}$O(p,$\gamma$)$^{17}$F data of Refs.~\cite{Chow1975,Morlock1997}. The small uncertainties quoted in that work do not include any systematic uncertainties of the data.}
\footnotetext[6] {\footnotesize Average result of R-matrix and potential model fits to the $^{16}$O(p,$\gamma$)$^{17}$F data of Refs.~\cite{Chow1975,Morlock1997}.}
\footnotetext[7] {\footnotesize Results of R-matrix fit to the $^{16}$O(p,$\gamma$)$^{17}$F data of Ref.~\cite{Morlock1997}. No uncertainties were reported.}
\end{center}
\end{table*}
\begin{table}[]
\begin{center}
\caption{Predicted total $S$ factors for $^{16}$O(p,$\gamma$)$^{17}$F. The uncertainty of the $S$ factor is $4.0$\% at all listed energies.}
\label{tab:results3}
\begin{ruledtabular}
\begin{tabular}{cccc}
E  & S(E) & E &  S(E)    \\
(MeV) & (MeVb) & (MeV) & (MeVb)  \\  
\hline
0.002   &   0.01081   &   0.100  &   0.00767    \\
0.004   &   0.01069   &   0.200  &   0.00639    \\
0.006   &   0.01059   &   0.400  &   0.00521    \\
0.008   &   0.01048   &   0.600  &   0.00465    \\
0.010   &   0.01038   &   0.800  &   0.00435    \\
0.020   &   0.00992   &   1.000  &   0.00413    \\
0.040   &   0.00915   &   1.500  &   0.00376    \\
0.060   &   0.00856   &   2.000  &   0.00356    \\
0.080   &   0.00807   &   2.500  &   0.00342    \\
\end{tabular}
\end{ruledtabular}
\end{center}
\end{table}

\begin{figure}[ht]
\includegraphics[width=1.0\linewidth]{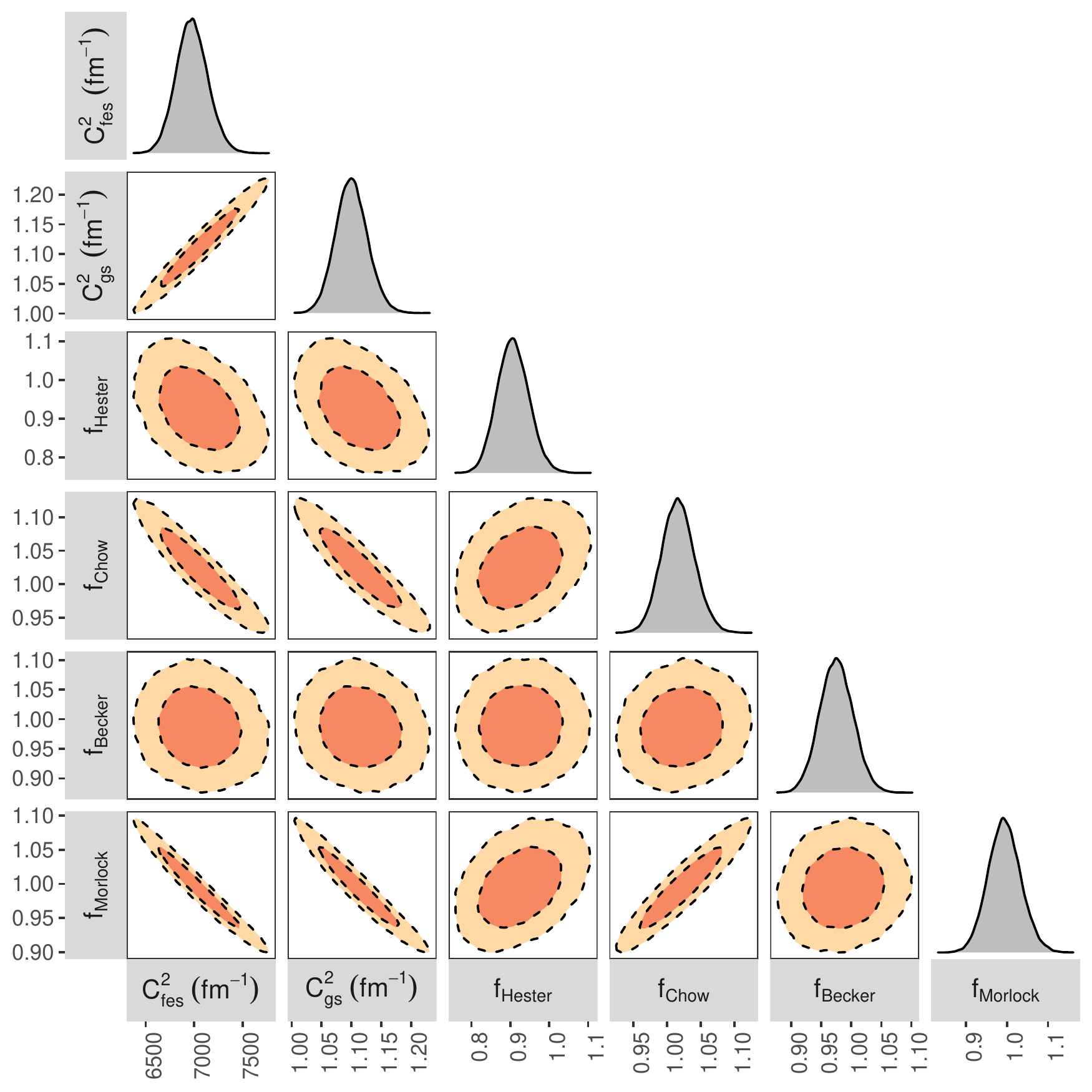}
\caption{\label{fig:posterior} 
One- and two-dimensional projections of the posterior probability distributions of two physical model parameters, i.e., the ANCs, $C^2_{gs}$ and $C^2_{fes}$, and four data set parameters, $f_j$, when adopting broad priors in the Bayesian fit. These results correspond to those given in column 2 of Table~\ref{tab:results1} (boldface) and in Table~\ref{tab:results2}). The dark and light shaded areas correspond to 68\% and 95\% coverage probabilities, respectively.}
\end{figure}

Posteriors of the two ANCs, $C^2_{gs}$ and $C^2_{fes}$, including their pairwise correlation, are presented in Fig.~\ref{fig:posterior}. As expected, the two quantities are highly correlated: when one is increased, the other one decreases to fit the total experimental $S$ factor. Our best-fit values are $C^2_{gs}$ $=$ $1.115$~fm$^{-1}$~($\pm$ 4.0\%) and $C^2_{fes}$ $=$ $7063$~fm$^{-1}$~($\pm$ 4.0\%). The results agree with those of Gagliardi et al. \cite{Gagliardi1999}. It is not surprising that their uncertainties ($\approx 10$\%) are significantly larger than ours, because their distorted wave Born approximation (DWBA) analysis of $^{16}$O($^3$He,d)$^{17}$F data was subject to ambiguities in the choice of optical model potentials for the incoming and outgoing channels. Our values are also in the vicinity of previously reported theoretical \cite{baye1998} and experimental results \cite{iliadis2008,Azuma2010}. A more detailed comparison is not possible, as no rigorous uncertainties were presented in these works. We do not present posteriors of the Woods-Saxon radius parameter, $r_0$, or the diffuseness, $a$, because the fit does not constrain these parameters, as discussed in Sec.~\ref{sec:nuclear}. 

Posteriors of the data-set normalization factors, $f$, are also depicted in Fig.~\ref{fig:posterior}. Numerical values for both $f.u.$ and $f$ are given in Table~\ref{tab:results2}. The former determine the spreads of the lognormal priors used to implement systematic uncertainties into the Bayesian model, while the latter are derived from the posteriors and represent our best estimate for the deviations between the data of a given experiment and the best-fit line. It can be seen that the $f$ values for the data of Refs.~\cite{Chow1975,Becker1982,Morlock1997} are consistent with unity for a coverage probability of 64\%. For the data of Ref.~\cite{Hester1958}, the $f$ value is consistent with unity for a coverage probability of 80\%. Overall, our Bayesian model predicts that the reported systematic uncertainties are consistent with the best-fit solution, i.e., we find no evidence for significant unreported systematic effects.
\begin{table}[]
\begin{center}
\caption{Normalization factors, $f$, of $^{16}$O(p,$\gamma$)$^{17}$F data sets.}
\label{tab:results2}
\begin{ruledtabular}
\begin{tabular}{lcc}
Experiment                                & $f.u.$\footnotemark[1] & $f$\footnotemark[2]    \\
\hline
Hester, Pixley \& Lamb \cite{Hester1958}  & $1.14$ &  $0.924 \pm 0.070$     \\
Chow, Griffiths \& Hall \cite{Chow1975}   & $1.05$ &  $1.022 \pm 0.041$     \\
Becker et al. \cite{Becker1982}           & $1.05$ &  $0.986 \pm 0.045$     \\
Morlock et al. \cite{Morlock1997}         & $1.10$ &  $0.994 \pm 0.041$     \\
\end{tabular}
\end{ruledtabular}
\footnotetext[1] {\footnotesize Systematic factor uncertainties of the experimental data (see Sec.~\ref{sec:data}). The values of $f.u.$ determine the spread of the lognormal probability density adopted for the {\it priors}. For example, a value of ``1.14'' represents a systematic uncertainty of ``14\%''.}
\footnotetext[2] {\footnotesize Results correspond to those given in boldface in column 2 of Table~\ref{tab:results1}, i.e., by assuming broad priors. The uncertainties correspond to 68\% coverage probabilities of the {\it posterior} densities.}
\end{center}
\end{table}

The small uncertainty of $4.0$\% in our results (column 2 of Table~\ref{tab:results1}), despite the fact that we varied the Woods-Saxon potential parameters over broad ranges, reflects the insensitivity of the ANCs, $C^2_{\ell_f}$, and fitted $S$ factor to the choice of potential parameters, as discussed in Sec.~\ref{sec:nuclear} and previously demonstrated by Gagliardi et al. \cite{Gagliardi1999} in the analysis of $^{16}$O($^3$He,d)$^{17}$F data. As an additional test of this assertion, we repeated the calculation by fixing the bound state potential parameters at values of $r_0$ $=$ $1.25$~fm and $a$ $=$ $0.65$~fm. The results are listed in column 3 of Table~\ref{tab:results1}. It can be seen that the values and their uncertainties obtained for the ANCs and S factors overall agree with those obtained by sampling over broad ranges of the potential parameters.

We will now turn to the discussion of spectroscopic factors. If we assume the conventional Woods-Saxon potential parameter values of $r_0$ $=$ $1.25$~fm and $a$ $=$ $0.65$~fm, we find $b^2_{gs}$ $=$ $0.8840$~fm$^{-1}$ and $b^2_{fes}$ $=$ $6899$~fm$^{-1}$ for the single-particle ANCs. From Eq.~(\ref{eq:relation}) and the results listed in column 2 of Table~\ref{tab:results1}, we obtain for the spectroscopic factors $S_{\ell_f = 2}$ $=$ $1.26\pm0.05$ and $S_{\ell_f = 0}$ $=$ $1.02\pm0.04$ for the transitions to the ground and first-excited state, respectively. These agree with previous theoretical and experimental analyses (see, e.g., Refs.~\cite{Vernotte1994,iliadis2004,iliadis2008}).

When the fit is performed using spectroscopic factors instead of ANCs as scaling factors (see Eq.~(\ref{eq:theory}), the obtained uncertainties become significantly larger. Assuming again broad priors for $r_0$ and $a$ (see Eqs.~(\ref{eq:prior1}) and (\ref{eq:prior1b})), the zero-energy $S$-factor uncertainty amounts to 9\% (i.e., more than twice the value discussed above). Similarly, the uncertainty of the derived spectroscopic factors increases substantially (to 11\%). These uncertainties can only be reduced if the sampling ranges of the priors for the potential parameters are significantly reduced. Generally, however, these parameters are poorly constrained by experiment and the commonly employed values of $r_0$ $=$ $1.25$~fm and $a$ $=$ $0.65$~fm are mainly adopted by convention. Clearly, performing the fits in terms of ANCs has a distinct advantage over using spectroscopic factors as scaling parameters.

\section{\label{sec:rates}Thermonuclear reaction rates}
The thermonuclear reaction rate per particle pair, $N_A \langle \sigma v \rangle$, at a given stellar temperature, $T$, is given by \citep{Iliadis:2015ta}
\begin{equation}
\begin{split}
N_A \langle \sigma v \rangle & = \left(\frac{8}{\pi \mu}\right)^{1/2} \frac{N_A}{(kT)^{3/2}}  \\ 
&  \times \int_0^\infty e^{-2\pi\eta}\,S(E)\,e^{-E/kT}\,dE, 
\label{eq:rate}
\end{split}
\end{equation}
where $\mu$ is the reduced mass of projectile and target, $N_A$ is the Avogadro constant, $k$ is the Boltzmann constant, and $E$ is the $^{16}$O $+$ $p$ center-of-mass energy.

The $^{16}$O(p,$\gamma$)$^{17}$F reaction rates were computed by integrating Equation~(\ref{eq:rate}) numerically. The $S$ factor is calculated from our MCMC samples (Sec.~\ref{sec:results}) and, therefore, the values of $N_A \langle \sigma v \rangle$ contain the effects of statistical and systematic uncertainties, as well as all correlations among parameters. The results are based on 5,000 random S-factor samples, which ensures that Markov chain Monte Carlo fluctuations are negligible compared to the reaction rate uncertainties. Our lower and upper integration limits were set to $0.002$~MeV and $2.5$~MeV, respectively. Reaction rates are computed for $50$ temperature grid points between $0.003$~GK and $3.5$~GK and numerical values presented in Table~\ref{tab:rates}.  The recommended rates are computed from the 50th percentile of the rate probability density function, while the factor uncertainty, $f.u.$, is obtained from the 16th and 84th percentiles \citep{Longland:2010is}. The new total rate uncertainties are 4.2\% for the entire temperature range shown. Notice that this uncertainty is slightly larger than that of the $S$ factor given in Table~\ref{tab:results3} (4.0\%). As already pointed out, the reason is that we computed the rates directly using the MCMC samples and, therefore, our results contain the information of parameter correlations.
%
\begin{table}[ht!] 
\begin{threeparttable}
\caption{Total thermonuclear reaction rates for $^{16}$O(p,$\gamma$)$^{17}$F \tnote{a}}
\setlength{\tabcolsep}{6pt}
\center
\begin{tabular}{ccccc}
\toprule
T (GK) & Low & Median &   High  &   f.u.  \\
\colrule 
 0.003 &  4.164E-41 &  4.338E-41 &  4.517E-41  &  1.042 \\
 0.004 &  1.343E-36 &  1.399E-36 &  1.457E-36  &  1.042 \\
 0.005 &  2.154E-33 &  2.244E-33 &  2.336E-33  &  1.042 \\
 0.006 &  5.980E-31 &  6.230E-31 &  6.486E-31  &  1.042 \\
 0.007 &  5.332E-29 &  5.555E-29 &  5.783E-29  &  1.042 \\
 0.008 &  2.162E-27 &  2.253E-27 &  2.345E-27  &  1.042 \\
 0.009 &  4.935E-26 &  5.142E-26 &  5.353E-26  &  1.042 \\
 0.010 &  7.290E-25 &  7.595E-25 &  7.907E-25  &  1.042 \\
 0.011 &  7.668E-24 &  7.988E-24 &  8.317E-24  &  1.042 \\
 0.012 &  6.148E-23 &  6.405E-23 &  6.669E-23  &  1.042 \\
 0.013 &  3.951E-22 &  4.117E-22 &  4.286E-22  &  1.042 \\
 0.014 &  2.114E-21 &  2.203E-21 &  2.293E-21  &  1.042 \\
 0.015 &  9.699E-21 &  1.010E-20 &  1.052E-20  &  1.042 \\
 0.016 &  3.903E-20 &  4.067E-20 &  4.234E-20  &  1.042 \\
 0.018 &  4.578E-19 &  4.770E-19 &  4.966E-19  &  1.042 \\
 0.020 &  3.809E-18 &  3.968E-18 &  4.131E-18  &  1.042 \\
 0.025 &  2.639E-16 &  2.749E-16 &  2.862E-16  &  1.042 \\
 0.030 &  6.641E-15 &  6.919E-15 &  7.204E-15  &  1.042 \\
 0.040 &  7.211E-13 &  7.513E-13 &  7.822E-13  &  1.042 \\
 0.050 &  1.998E-11 &  2.082E-11 &  2.167E-11  &  1.042 \\
 0.060 &  2.497E-10 &  2.601E-10 &  2.708E-10  &  1.042 \\
 0.070 &  1.865E-09 &  1.943E-09 &  2.023E-09  &  1.042 \\
 0.080 &  9.751E-09 &  1.016E-08 &  1.058E-08  &  1.042 \\
 0.090 &  3.933E-08 &  4.098E-08 &  4.267E-08  &  1.042 \\
 0.100 &  1.304E-07 &  1.359E-07 &  1.414E-07  &  1.042 \\
 0.110 &  3.709E-07 &  3.865E-07 &  4.024E-07  &  1.042 \\
 0.120 &  9.340E-07 &  9.731E-07 &  1.013E-06  &  1.042 \\
 0.130 &  2.129E-06 &  2.219E-06 &  2.310E-06  &  1.042 \\
 0.140 &  4.472E-06 &  4.659E-06 &  4.851E-06  &  1.042 \\
 0.150 &  8.764E-06 &  9.131E-06 &  9.507E-06  &  1.042 \\
 0.160 &  1.620E-05 &  1.688E-05 &  1.757E-05  &  1.042 \\
 0.180 &  4.791E-05 &  4.991E-05 &  5.197E-05  &  1.042 \\
 0.200 &  1.216E-04 &  1.266E-04 &  1.319E-04  &  1.042 \\
 0.250 &  7.781E-04 &  8.106E-04 &  8.440E-04  &  1.042 \\
 0.300 &  3.176E-03 &  3.309E-03 &  3.446E-03  &  1.042 \\
 0.350 &  9.703E-03 &  1.011E-02 &  1.053E-02  &  1.042 \\
 0.400 &  2.426E-02 &  2.528E-02 &  2.632E-02  &  1.042 \\
 0.450 &  5.244E-02 &  5.464E-02 &  5.689E-02  &  1.042 \\
 0.500 &  1.016E-01 &  1.058E-01 &  1.102E-01  &  1.042 \\
 0.600 &  3.000E-01 &  3.126E-01 &  3.255E-01  &  1.042 \\
 0.700 &  7.083E-01 &  7.381E-01 &  7.683E-01  &  1.042 \\
 0.800 &  1.432E+00 &  1.493E+00 &  1.554E+00  &  1.042 \\
 0.900 &  2.590E+00 &  2.699E+00 &  2.809E+00  &  1.042 \\
 1.000 &  4.301E+00 &  4.482E+00 &  4.666E+00  &  1.042 \\
 1.250 &  1.180E+01 &  1.229E+01 &  1.280E+01  &  1.042 \\
 1.500 &  2.526E+01 &  2.632E+01 &  2.740E+01  &  1.042 \\
 1.750 &  4.612E+01 &  4.806E+01 &  5.005E+01  &  1.042 \\
 2.000 &  7.545E+01 &  7.863E+01 &  8.188E+01  &  1.042 \\
 2.500 &  1.618E+02 &  1.686E+02 &  1.755E+02  &  1.042 \\
 3.000 &  2.851E+02 &  2.971E+02 &  3.093E+02  &  1.042 \\
 3.500 &  4.406E+02 &  4.591E+02 &  4.779E+02  &  1.042 \\
\botrule \\
\end{tabular}
\begin{tablenotes}
\item[a] {\footnotesize In units of cm$^3$mol$^{-1}$s$^{-1}$. Columns 2, 3, and 4 list the 16th, 50th, and 84th percentiles, respectively, of the total rate probability density function at the given temperatures; $f.u.$ is the factor uncertainty of the total reaction rate, based on Monte Carlo sampling. The total number of samples at each temperature was 5,000.}
\end{tablenotes}
\label{tab:rates}
\end{threeparttable}
\end{table}    

\begin{figure}
\includegraphics[width=1\linewidth]{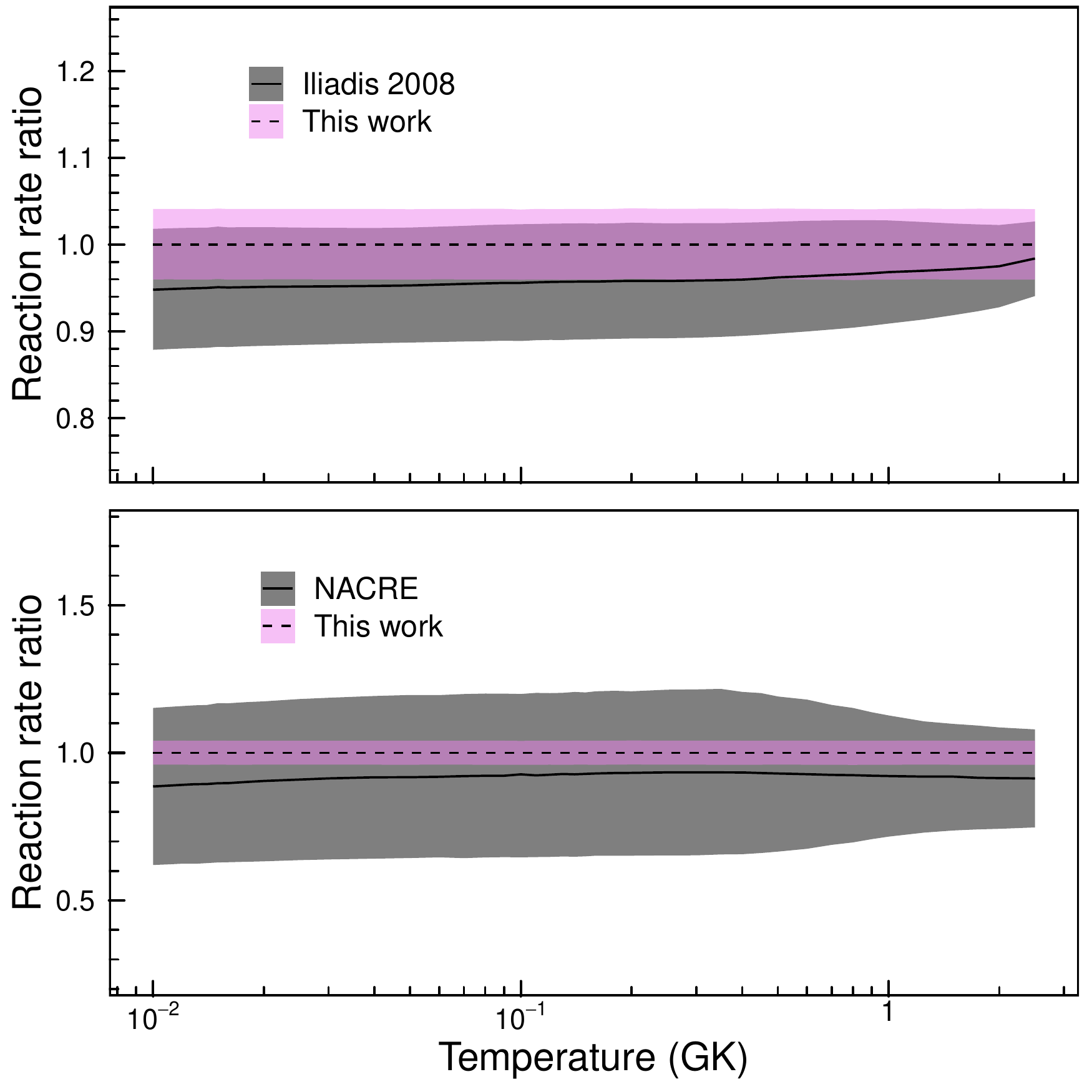}
\caption{Comparison of present to previously evaluated $^{16}$O(p,$\gamma$)$^{17}$F reaction rates. For better comparison, all rates have been normalized to the present median rate (see column 3 in Table~\ref{tab:rates}). The magenta-shaded regions correspond to our 68\% uncertainties (see column 5 in Table~\ref{tab:rates}). The gray-shaded regions depict the rates and uncertainties reported in Ref.~\cite{iliadis2008} (top panel) and Ref.~\cite{NACRE} (bottom panel).}
\label{fig:rates}
\end{figure}

The new rates are displayed in Fig.~\ref{fig:rates} and compared to previous results. For better comparison, all rates have been divided by the present median (50th percentile) values (column 3 of Table~\ref{tab:rates}). Our low (16th percentile) and high (84th percentile) rates are depicted as magenta-shaded bands centered around unity (dashed line). The gray bands refer to the rates of \citet{iliadis2008} (top panel) and \citet{NACRE} (bottom panel). It can be seen that our median rate is about 5\% and 8\% higher than those of \citet{iliadis2008} and \citet{NACRE}, respectively. Furthermore, the uncertainties of our new rate ($4.2$\%) are about half of those reported by Ref.~\cite{iliadis2008} ($7.5$\%) and about a factor of $8$ smaller than those of Ref.~\cite{NACRE} ($30$\%). As already pointed out above, all previously evaluated rates lack a rigorous statistical interpretation based on coverage probabilities.  

\section{\label{sec:summary}Summary}
We presented the first statistically rigorous evaluation of the $^{16}$O(p,$\gamma$)$^{17}$F direct capture $S$ factor and thermonuclear reaction rates. Our analysis includes four data sets that reported statistical and systematic uncertainties separately \cite{Chow1975,Becker1982,Morlock1997}. The combined fit for the transitions to the ground and first-excited states in $^{17}$F, and their sum, was performed using a Bayesian method. The physical model was a single-particle model employing a Woods-Saxon potential for generating the radial bound state wave function. The fit had three adjustable parameters: the radius parameter and diffuseness of the Woods-Saxon potential, and the asymptotic normalization coeffients (ANCs) for scaling the theoretical direct capture $S$ factor. We found that a poor fit is obtained when it is performed using spectroscopic factors as scaling parameters instead. Since the $^{16}$O(p,$\gamma$)$^{17}$F reaction at low bombarding energies is of peripheral nature, the analysis of the $S$ factor in terms of ANCs greatly reduces the sensitivity to the single-particle potential parameters. For the ANC's of the ground and first-excited state transitions, we find values of $C^2_{gs}$ $=$ $1.115$~fm$^{-1}$ ($\pm 4.0$\%) and $C^2_{fes}$ $=$ $7063$~fm$^{-1}$ ($\pm 4.0$\%), respectively. The total $S$ factor at zero energy is $S_{gs+fes}(0)$ $=$ $0.01092$~MeVb ($\pm 4.0$\%). The thermonuclear rate is computed from the Markov Chain Monte Carlo (MCMC) samples by numerical integration and, therefore, accounts for all correlations between parameters. The rate uncertainties between $3$~MK and $3.5$~GK is $4.2$\%, about half the previously reported rates \cite{iliadis2008}.

Our recommendation for future work is to remeasure the $^{16}$O(p,$\gamma$)$^{17}$F reaction cross section at center-of-mass energies below $200$~keV, which have so far only been reached in the experiment of Hester, Pixley and Lamb \cite{Hester1958}, albeit with a relatively large systematic uncertainty of $14$\%. Finally, we emphasize that the Bayesian fit results presented here are based on a specific nuclear reaction (i.e., a single-particle potential) model. It would be interesting to compare these results to those from a future R-matrix analysis of the same data as analyzed here.

\begin{acknowledgments}
We would like to thank Art Champagne and Richard Longland for helping with calculating the Whittaker function. We also would like to express our gratitude Jo Moscoso for his help during the early stages of this project. The constructive comments of Robert Janssens are highly appreciated. This work was supported in part by the DOE, Office of Science, Office of Nuclear Physics, under grants DE-FG02-97ER41041 (UNC) and DE-FG02-97ER41033 (TUNL).\end{acknowledgments}



%

\end{document}